\documentclass[10pt,twocolumn,conference]{IEEEtran}
\makeatletter
\def\ps@headings{%
\def\@oddhead{\mbox{}\scriptsize\rightmark \hfil \thepage}%
\def\@evenhead{\scriptsize\thepage \hfil \leftmark\mbox{}}%
\def\@oddfoot{}%
\def\@evenfoot{}}
\makeatother 

\usepackage{setspace}
\usepackage[final]{graphicx}
\usepackage[reqno]{amsmath}
\usepackage{amssymb}
\usepackage{cite}
\usepackage{balance}
\usepackage{color}
\usepackage{subfigure}
\usepackage[inoutnumbered,linesnumbered,algoruled,slide,vlined]{algorithm2e}
\usepackage{algorithmic}

\newfont{\bbb}{msbm10 scaled 500}

\newfont{\bb}{msbm10 scaled 1100}

\newcommand{\argmax}{\operatornamewithlimits{argmax}}
\newcommand{\argmin}{\operatornamewithlimits{argmin}}

\newtheorem{lemma}{Lemma}%[chapter]
\begin{document}

    \title{Learning based Channel Load Measurement in 802.11 Networks }
    \author{\IEEEauthorblockN{Mehmet Karaca and Bj{\"o}rn Landfeldt \\}
    \IEEEauthorblockA{The Department of Electrical and Information Technologies, Lund University, Lund/Sweden. \\Email:
    mehmet.karaca@eit.lth.se, zhi.zhang@eit.lth.se, bjorn.landfeldt@eit.lth.se}\thanks{This  work  was  partially  sponsored  by  the  EC  FP7  Marie Curie IAPP Project 324515, "MeshWise"}}

\maketitle

%\title{Dynamic Control of Wireless Networks with Secrecy}
\begin{abstract}
It is known that load-unaware channel selection in 802.11 networks results in high interference, which can significantly reduce network throughput. In current implementations, the only way to determine traffic load on a channel is to measure the channel. Therefore, in order to find the channel with the minimum load, all channels have to be measured, which is costly and may cause unacceptable communication interruptions between the Access Point (AP) and the
stations (STA). In this paper, we propose a learning based approach which seeks the channel with the minimum load by measuring only a limited number of channels. Our method uses Gaussian Process Regressing to accurately track the traffic load on each channel based on  previous measured load.  We confirm the performance of our approach using experimental data, and
show that the time used for the load measurement can be reduced more than  50 $\%$
compared to the case where all channels are monitored.
\end{abstract}

\section{Introduction}
%IEEE 802.11 has become the de-facto standard for WLANs which have been widely deployed in various areas. The  acceptance of the IEEE 802.11 protocol is expected to become wider  since 
It is estimated that more than 60\% of global Internet traffic will be transmitted over IEEE 802.11 based Wireless Local Area Networks (WLANs) in 2018 \cite{Costa}, bringing high density unstructured networks. Hence, industry efforts such as the IEEE 802.11ac standard are focussed on very high throughput. Moreover the High-Efficiency WLAN (HEW) Study Group \cite{IEEE11ax} is currently working on a new high-throughput amendment named IEEE 802.11ax-2019 which aims to improve  user experience  especially in dense deployment scenarios. Under such scenarios with high interference levels,  identifying the channel with the least traffic load is crucial.

%IEEE 802.11-based WLANs~\cite{IEEE:80211} have been widely deployed
%at airports, schools, homes etc. This rapid growth in the use of
%WLAN products has caused huge amount of interference and contention
%among the deployed APs. As a result,  the development of efficient
%radio resource management algorithms has been emerged. One of the
%most important properties of such an algorithm is to maximize the
%network throughput while providing some level of Quality of Service
%(QoS) to each station. To realize this property, each AP must
%operate on the \emph{best} channel with the minimum interference and
%traffic load. However, APs must be aware of the levels of
%interference and traffic load on each available channel in the
%interested frequency band so that the best channel can be selected.

In practical implementations, the traffic load on a particular channel (i.e., channel load) is measured using the Clear Channel Assessment (CCA) mechanism which can measure the fraction of time in which the channel is busy or idle~\cite{CCA}.  Acquisition  of  the channel load information has been standardized with 
the IEEE 802.11k standard~\cite{IEEE:80211k}, where
measurements are performed with request/response frame exchanges. Specifically, by sending a \emph{channel load request} frame,  an AP or STA can request another AP or STA to measure the load of a particular set of channels using CCA. Then, the station that measured the channels returns the
channel busy fraction on those channels by sending a \emph{channel load report} frame.

We note that CCA based load measurement may take significant time since the monitoring station should halt its transmission/reception for the duration of the measurement. To give an idea of how much time is needed to collect load information of each channel,  we consider the 5 GHz frequency band where there are 23 non-overlapping channels with a bandwidth of 20 MHz. If a channel is monitored for a duration of 50
milliseconds (ms) then the total time spent for the monitoring
process will equal 1150 ms (1.15 seconds), which can significantly degrade the performance of the monitoring station in terms of both the throughput and delay. If the monitoring station is the AP, then the effect of the monitoring becomes more significant.

One way to reduce the overhead of the load measurement is to decrease the measurement time. However, 
the confidence of each measurement is important parameter since a
channel is measured for only a limited time. In~\cite{80211k2}, it
was shown that the measurement duration must be sufficiently large
for a certain level of confidence to be guaranteed.
In~\cite{80211k1}, the authors studied the optimization of the
duration of a single load measurement. It was shown
in~\cite{80211k3} that there is significant variation in channel
loads reported by the same station at different times, which may
have significant effect on the selection of the channel with the
minimum load. In~\cite{Tassiulas}, the authors proposed a channel selection mechanism which takes into account the channel load without considering the cost of obtaining the load information.

Another approach to reducing the time spent on load measurement is to monitor only a limited number of channels at each measurement time instead of monitoring all channels, which is the approach considered in this paper. Specifically, we propose a dynamic
load acquisition algorithm which aims to determine the channel with
the minimum traffic load without measuring all channels in the frequency band of interest. Our algorithm is based on the Gaussian
Process Regression (GPR) technique~\cite{Rasmussen:GP}, which is
used to estimate the instantaneous load of each channel by utilizing  the previous load measurements. Based on the estimated load and the level of uncertainty in the estimations, it constructs a set of channels to be measured, and only those channels are measured at each measurement time. We show that GPR-based load measurement
works well for reducing the cost associated with channel
monitoring.

%Our contributions are summarized as follows:
%%\subsection{Contributions}
%\begin{itemize}
%\item Based on the work in~\cite{Alpcan:Valuetools11}, Gaussian Process Regression learning algorithm is proposed to track the channel load.
%\item A joint scheduling and probing algorithm is proposed in which the subset of users
%probed at every slot is adaptively selected based on the dynamics of
%the channel processes.
%\item We implement a realistic network setting where we simulate High Data Rate (HDR) protocol in CDMA cellular
%networks, and wireless channel is modeled as time-correlated and
%non-stationary.  We show by numerical analysis that when our
%proposed algorithm is used the network can carry higher user traffic
%compared to Max-Weight algorithm with full CSI.
%\end{itemize}

\section{System Model}
We first describe our testbed to collect traffic load experimentally. 
In our testbed, we use  a  Wi-Fi device as a measuring station equipped with a Broadcom 802.11n chipset. Although the card does not support 802.11k,  it still enables us to measure load\footnote{ Traffic load on a channel is
caused by not only the transmission of data packets but also the
transmission of other management and control packets (i.e., beacons,
RTS/CTS, ACK packets).} using the CCA mechanism, and we examine the traffic load from the \textit{wireless driver} of the device. We recall that CCA is a function which 
senses the wireless medium. The channel measurement period is denoted $T$. We note that $T$ depends on the algorithm implemented on the device, and can be modified by end-users. In practice, the channel measurement is performed in a discrete way. Specifically,  $T$ is divided into mini-slots of fixed duration, which cannot be changed by end-users (i.e., depends on  the card clock). The CCA mechanism returns a 1 if the channel is busy during that mini-slot and otherwise it returns a 0. Let $n(T)$ be the number of mini-slots when the measurement duration is set to $T$. Then, the fraction of busy time of a channel is determined by averaging the results obtained with $n(T)$ samples. As $T$ increases,  the number of samples (i.e., mini-slots) increase as well, and the measurements become more accurate.

%
%Operating on a crowded channel in either frequency band degrades the
%performance of the network significantly in terms of throughput and
%delay. Hence, before deciding the operating  channel, an AP must
%obtain traffic load information of each channel. 802.11k standard
%provides a simple way for APs to gather this information, which is
%realized as follow: first the AP asks a measuring station (MS) to
%measure the traffic load\footnote{ Traffic load on a channel is
%caused  by not only the transmission of data packets but also the
%transmission of other management and control packets (i.e., beacons,
%RTS/CTS, ACK packets).}  on a set of channels for a duration of $T$
%seconds. Then, the MS reports this information back to the AP. In
%some cases, a station might request information from APs. In this
%work, we assume that the AP asks a single MS to report channel load
%information.

\subsection{An Exhaustive Algorithm}
Let $\mathcal{C}_a=\{C_1,C_2,\cdots,C_{M}\}$ be the set of all
available channels in the operating frequency band, and
$L(t)=\{L_(t),L_2(t),\cdots,L_{M}(t)\}$ be the traffic load 
vector where $L_n(t)$ represents the
traffic load on channel
$n$  when the measurements starts at time $t$. We assume that each channel is measured for a duration of $T$ seconds. To gather the load information from all channels,  Algorithm 1 is applied, which is an exhaustive algorithm that measures all the available channels in the frequency band of interest.
\begin{algorithm}
\begin{itemize}
\item Step 1: The monitoring station (MS)  receives the measurement request for a
duration of  $T$ seconds. The request may come from another station or AP. 

\item Step 2: When the MS starts the measurements:

\begin{itemize}
\item Step 2.1: First, the MS halts its transmission/reception, and measures the load on each
channel in $\mathcal{C}_a$  for $T$ seconds using CCA.

\item Step 2.2: Then, the MS reports this information back to the AP.
\end{itemize}

\item Step 3: Let $ k_1^*(t)$ be the channel with the minimum load at measurement time $t$.
\begin{align*}
k_1^*(t)=\argmin_{n \in \mathcal{C}_a} L_n(t)
\end{align*}
%\item Step 3: If $m(t) \neq k_1^*(t)$, then the AP switches to channel
%$k_1^*(t)$. Otherwise it continues operating on channel $m(t)$.
\end{itemize}
\caption{\small{Exhaustive Load Measurement}}
\label{tauupdate}
\end{algorithm}

Let $L_{k_1^*(t)}$ be the load on channel $k_1^*(t)$ selected by
Algorithm 1 at measurement time $t$. Then, the average channel load
using Algorithm 1 is given as,
\begin{align}
L_1=\lim_{t\rightarrow \infty}\frac{1}{t} \sum_{\tau=0}^{t-1}
L_{k_1^*(\tau)} \label{eq:avgload1}
\end{align}
Let $c_1$ be the cost in terms of
time consumed with Algorithm 1.  Since all the channels are measured by Algorithm 1,  $c_1=T  \times M$.

The cost of employing Algorithm 1 is non-negligible since it
requires the MS to monitor a large number of channels for a
non-negligible duration. We recall that the following options are available to
reduce the time spent for the measurement process: i) 
decrease $T$, and then the overall time spent for monitoring all
channels will be reduced as well. However, as we show in our
experimental results given in Section IV, channel selection with
small values of $T$ may result in incorrect decisions, and the
required confidence level may not be satisfied~\cite{80211k2}; ii-)
measure only a channel subset instead of all channels.

In our work, we adopt the second solution, and consider that at most
$K$ channels can be monitored at a measurement request, where $K ²
M$. We define $\mathcal{C}_l(t)$ as the set of channels monitored at
measurement time $t$, where $\mid \mathcal{C}_l(t) \mid =K$ for all
$t$. Recall that  when all channels are measured as in Algorithm 1,
the channel with the minimum load is guaranteed to be selected.
However, when $K$ channels are monitored, it is not guaranteed to
find the channel with the minimum traffic load. Hence, it is
important to determine the set of channels that should be monitored.
Note that the instantaneous measured data may be outdated at the
time of channel selection due to the fast variation of the load
processes. By taking this into account, in this work we adopt
an estimation based solution for the determination of the set of
channels, where we predict the instantaneous average of the load
process at each measurement time. 

\section{Channel Load Estimation with GPR}
We employ GPR for channel load
estimation~\cite{Rasmussen:GP}. GPR is a popular learning method for
predicting and tracking of continuous processes, and it is widely
used especially for practical problems including global
optimization~\cite{Alpcan:Valuetools11}, wireless
scheduling~\cite{karacaICC}, global positioning~\cite{GP:positioning}
and estimation in wireless sensor networks~\cite{Gu12}. Note
that the foundation of GPR is Bayesian
inference, where the main idea is to choose an a priori model and
update this model with observed measurements. GPR is a suitable approach for the following reasons; i-) GPR is a nonparametric
regression model, and the current state of the underlying process
can be estimated using only some previous measurement samples; ii-)
GPR provides a simple way to measure the uncertainty in the
estimation for any given set of channel load measurements. This is
particularly important for systems where only limited measurement data exists. ; ii-) Note that the channel load process
may be highly non-stationary. GPR can
give estimations for the current state of the process using only the most
recent measurement results, and this is especially important for
non-stationary processes, since previous measurements may become
outdated and may not give accurate information about the current state.

Recall that GPR aims to reconstruct the underlying function with limited data, which in our case is the traffic load process.  It is important to highlight that the performance of GPR highly depends on how smooth the underlying function is. From our experimental data, we observe that the difference between even two consecutive measurements can be very high, which prevents us from obtaining a smooth function for GPR to work well. In order to make the traffic load process smoother, we employ a linear smoother which uses the moving average by using the most recent $w$ instantaneous load measurements.  Specifically, let $\mathcal{D}_n(t)=(\textbf{L}_n^a,\boldsymbol \tau_n)$ denote the
set of channel load measurements taken in channel $n$ at the
beginning of measurement period $t$, where $\textbf{L}_n^a=\{L_{n_1}^a,
L_{n_2}^a, \dots, L_{n_w}^a\}$ denotes the set of the averaged traffic load using the latest  $w$ 
instantaneous channel load measurements at times, $\boldsymbol
\tau_n=\{\tau_n^1,\tau_n^2,\dots,\tau_n^w\}$, and $\tau_n^i < t$,
$\forall \tau_n^i \in \boldsymbol \tau_n$, $i\in \{1,2,\dots,w\}$.
\begin{align}
i^*=\argmax_{1\leq n\leq N} I_n(t)=\argmax_{1\leq n\leq N} v_n(t).
\label{eq:proposition}
\end{align}

Here, we define the \emph{instantaneous average load} of  a channel
at time $t$ as the sample average of the latest $w$ measurements
taken until time $t$, where the value of $w$ depends on how fast the
measurement statistics change on the channel. We use GPR to
determine $\hat{L}_n(t)$ given $\mathcal{D}_n(t)$ instead of simply
averaging the latest $w$ measurements.

The following lemma is similar to the one given
in~\cite{Alpcan:Valuetools11}, and establishes that the information
obtained by probing a channel is equal to the variance of the
estimate of the state of that channel.
\begin{lemma}
\label{prop:valuetools} Given $\mathcal{D}_n(t), \forall n=1,\ldots,
N$, finding the channel that gives the best information at time
slot $t$ is equal to finding the channel which has the highest
variance at that time slot, i.e.,
\begin{align}
i^*=\argmax_{1\leq n\leq N} I_n(t)=\argmax_{1\leq n\leq N} v_n(t).
\label{eq:proposition}
\end{align}
\end{lemma}

Let $p(L_n(t)| t,\mathcal{D}_n(t))$ be a posterior distribution of
channel $n$. According to GPR, a posterior distribution is Gaussian
with mean $\hat{L}_n(t)$ and variance $v_n(t)$. Specifically,
the Gaussian process is specified by the kernel function, $k_n(\tau_n^i,
\tau_n^j)$, that describes the correlation of the load on channel
$n$ between two measurements taken at times $\tau_n^i$ and
$\tau_n^j$. It is possible to choose any positive definite kernel
function. However, the most widely used is the squared exponential,
i.e., Gaussian, kernel:
\begin{align}
k_n(\tau_n^i,
\tau_n^j)=\exp\left[-\frac{1}{2}(\tau_n^i-\tau_n^j)^2\right].\label{eq:kernel1}
\end{align}
Given $\mathcal{D}_n(t)$,  $\hat{L}_n(t)$ and variance $v_n(t)$ are
determined as follows:
\begin{align}
\hat{L}_n(t)&=\textbf{k}_n^T(t)\textbf{K}_n^{-1}\textbf{L}_n,\label{eq:mean1}\\
v_n(t)&=k_n(t,t)-\textbf{k}_n^T(t)\textbf{K}_n^{-1}\textbf{k}_n(t),
\label{eq:var1}
\end{align}
where $\textbf{K}_n$ is a $w \times w$ matrix composed of elements
$k_n(\tau_n^i, \tau_n^j)$ for $1\leq i,j\leq w$ and
$\textbf{k}_n(t)$ is a vector with elements $k(\tau_n^i,t)$ for
$\forall \tau_n^i \in \boldsymbol \tau_n$. Hence, the AP can easily
predict the load on each channel at time $t$ by using
\eqref{eq:mean1}. Furthermore, the variance $v_n(t)$ is used to
measure the level of uncertainty in the estimation.

Note that an estimation cannot be done without some level of
uncertainty. The degree of uncertainty in the estimation of the
current process highly depends on the previously gathered
measurement data and the dynamics of the process. For instance, the
uncertainty level in the estimation of the current state of the
channel which was monitored recently is less than the channel which
has not been measured for a long time. Similar to the work
in~\cite{Alpcan:Valuetools11},  we use $v_n(t)$ as the degree of the
uncertainty in the estimation of the channel load. We have two
objectives; the first is to minimize the channel load. The second
 is to measure each channel closely and to acquire as much
information about the current load levels of the channels as
possible so that the estimation variance,$v_n(t)$, is minimized.
Next, we introduce our algorithm that aims to meet these two objectives
concurrently.

\subsection{Channel Selection Algorithm with GPR} \label{sec:algs} Here, we
propose our algorithm  that selects $K$ channels at every
measurement time.
\begin{algorithm}
\begin{itemize}
\item Step 1: The AP receives the latest $w$  load measurements for each
channel in $\mathcal{C}(t)$. Then, for each channel the AP:

\begin{itemize}
\item Step 2.1: Calculates $\hat{L}_n(t)$ and $v_n(t)$ according
to \eqref{eq:mean1} and \eqref{eq:var1}. The AP assigns a weight for
each channel as follows:
\begin{align*}
W_n(t)=v_n(t)\hat{L}_n(t)
\end{align*}

\item Step 2.2: Sorts $W_n(t)$ in descending order.

\item Step 2.3: Pick the first $K$ channels in the set denoted $\mathcal{C}_l(t)$ .
\end{itemize}

\item Step 3: Then, the AP requests the MS to monitor the
channels in set $\mathcal{C}_l(t)$.

\item Step 4: Let $m(t)$ be the current operating channel of the AP
when the load measurement is requested.
\begin{align*}
k_2^*(t)=\argmin_{n \in \mathcal{C}_l(t)} L_n(t)
\end{align*}
\item  Step 5: If $m(t) \neq k_2^*(t)$, the AP switches to
channel $k_2^*(t)$, otherwise it continues operating on channel
$m(t)$.
\end{itemize}
\caption{\small{Channel Load Measurement with GPR}}
\label{tauupdate}
\end{algorithm}
%Note that the only difference between Algorithm 1 and Algorithm 2 is
%that all channels are monitored by Algorithm 1 whereas only $K$
%channels are measured by Algorithm 2, and the channel selection is
%performed among  these $K$ channels.

Let $L_{k_2^*(t)}$ be the load on channel $k_2^*(t)$ selected by
Algorithm 2 at measurement time $t$. Then, the average channel load
with Algorithm 2 is given as,
\begin{align}
L_2=\lim_{t\rightarrow \infty}\frac{1}{t} \sum_{\tau=0}^{t-1}
L_{k_2^*(\tau)} \label{eq:avgload2}
\end{align}
Let $c_2$ be the cost in terms of time consumed for monitoring $K$
channels with Algorithm 2, and $c_2=T  \times K$.

\section{Simulation Results}
\label{sec:sim} In this section, we provide the results of the
impact of $T$ on channel selection, and of the performance
assessment of Algorithm 2 in terms of $L_1$, $L_2$, $c_1$ and $c_2$.
Our tests are carried out at the AirtTies office for the 2.4 GHz band
where there are 13 channels with 20 MHz bandwidth (i.e., $M=13$).

\subsection{Effect of Measurement duration, $T$}
In this part, we investigate the possible effects of the measurement
duration $T$ on the performance of a channel selection algorithm.
For this, we have conducted various channel load measurements for
different values of $T$.  During the measurements there were about 10
APs serving  more than 50 people.  The majority of the APs are of type
AirTies 4420 with Broadcom 4717 chipsets where the IEEE 802.11n
standard is supported.

The Broadcom 4717 chipsets do not support the request/response frames of
802.11k for  measurements, but they are capable of using CCA which
allows us to obtain the channel load information.

In our first test, the CCA supporting AP monitors each channel for a
duration of 10 ms to gather the load information. The measurements
are performed in a consecutive cyclic manner where the AP first measures
channel 1, then  channel 2 up to channel 13 and then immediately starts over for a total of 50 samples per channel.
\begin{figure}[t]
    \centering
    \includegraphics[width=1.0\columnwidth]{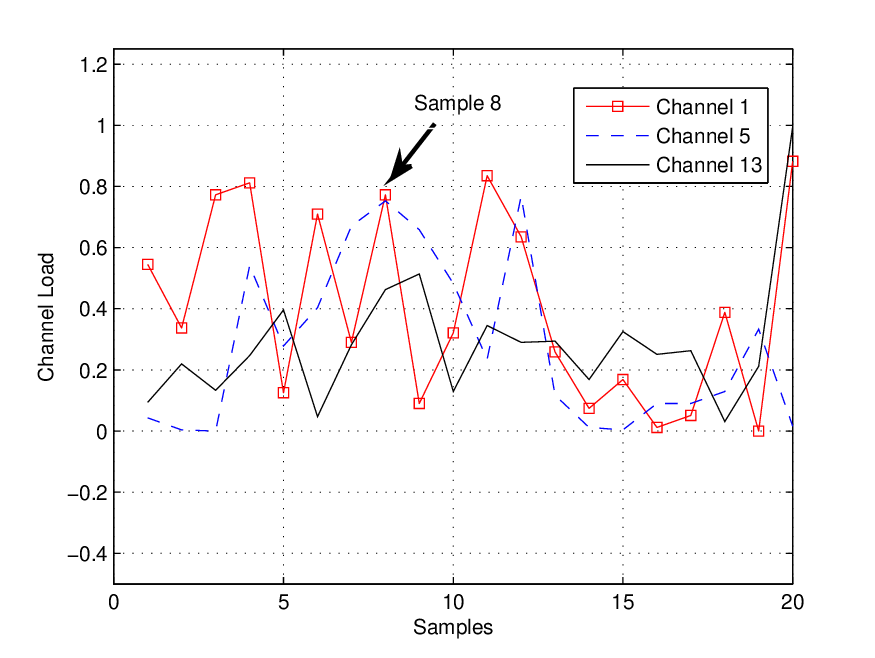}
    \caption{Channel load measurements when $T=10$ ms.}
    \label{fig:fig1}
\end{figure}
Figure 1 shows the channel load measured when $T=10$ ms. For clarity we only
plot the results for  channel 1 channel 5 and channel 13 as we observe
that the other channels show similar characteristics. It can be observed that the variations in the
traffic load  is high for all channels, which indicates that the
load process is non-stationary when $T=10$ ms. The effect of this
behavior of the load process leads to proneness for erroneous channel selection. For instance, in Figure 1 we
have highlighted sample 8. At this point if the AP monitors the spectrum for channel
selection, it observes that the load on channel 1 and channel 5 are
equal to 0.77 whereas it is equal to 0.46 for channel 13. Based on
this information, the AP decides to operate on channel 13. However, at
the next sample point the load on channel 1, channel 5 and channel
13 are 0.09, 0.66 and 0.51 respectively. Hence, at point 9 the best channel with the minimum load is channel 1 and
not channel 13. Also, monitoring each channel for an insufficient
duration may cause frequent  channel switching, which brings
additional costs in terms of increased switching delay and frequent user
disassociation. Hence, it is important to monitor each channel for a
duration large enough so that a sufficient number of samples,
$n(T)$, can be obtained, and the average of the obtained samples
gives accurate results.

Taking this into account, we repeat the experiment with
larger values of $T$. Figure 2 depicts the channel load gathered
when $T=50$ ms and $T=100$ ms. Clearly, as $T$ increases the load
curves smoothens out and the non-stationarity level decreases, thus, the problems associated with the
non-stationary nature of the load process can be mitigated. Next, we
present the performance of Algorithm 2 using the load data
gathered when $T=100$ ms.
\begin{figure}[t]
    \centering
    \includegraphics[width=1.0\columnwidth]{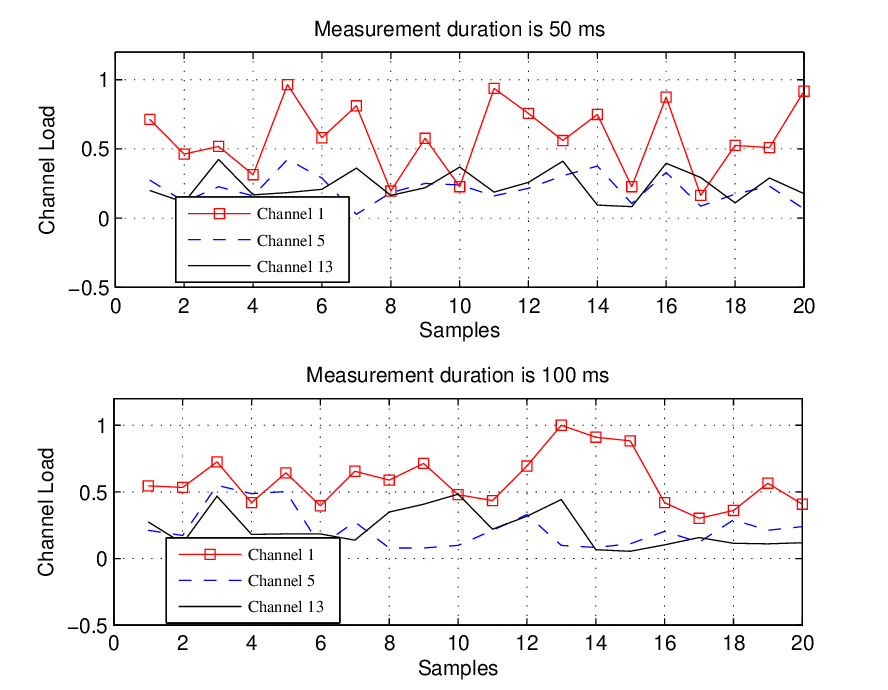}
    \caption{Channel load measurements when $T=50$ and $T=100$ ms.}
    \label{fig:fig2}
\end{figure}
\subsection{Performance of Algorithm 2}
In this part, we present the results of the performance assessment
of GPR in reducing measurement cost. For this, we apply Algorithm 2
and compare it with Algorithm 1 where all channels are monitored at
each measurement time. We also compare Algorithm 2 with a benchmark
algorithm which gives the estimation of the instantaneous average of
a channel load by averaging the latest $w$ measurement samples.

Figure 3 depicts $L_1$ and $L_2$ which are the channel load averaged
over 50 measurement points after Algorithm 1 and Algorithm 2 are
applied, respectively. For Algorithm 2, we change $K$ from $K=2$ to
$K=10$, and set $w=2$. Algorithm 1 always selects the channel with
the minimum load at each point, and the average channel load is
approximately 0.2163, i.e., $L_1=0.063$. Clearly, as $K$ increases,
the average estimated channel load using  Algorithm 2, $L_2$ decreases since
the accuracy of the estimation  increases with GPR, as it tracks
the load process well with larger values of $K$. When $K=7$, we
observe that $L_2$ is approximately equal to 0.065. This means
Algorithm 2 can achieve 96 $\%$  of the performance of Algorithm 1
by only monitoring $K=8$ channels. On the other hand, the monitoring
cost $c_1$ is equal to $13 \times 100 =1300$ ms whereas $c_2$ is
equal to $7 \times 100 =700$ ms. Hence, approximately 54 $\%$ of the
cost can be reduced using Algorithm 2 in this scenario.
\begin{figure}[t]
    \centering
    \includegraphics[width=1.0\columnwidth]{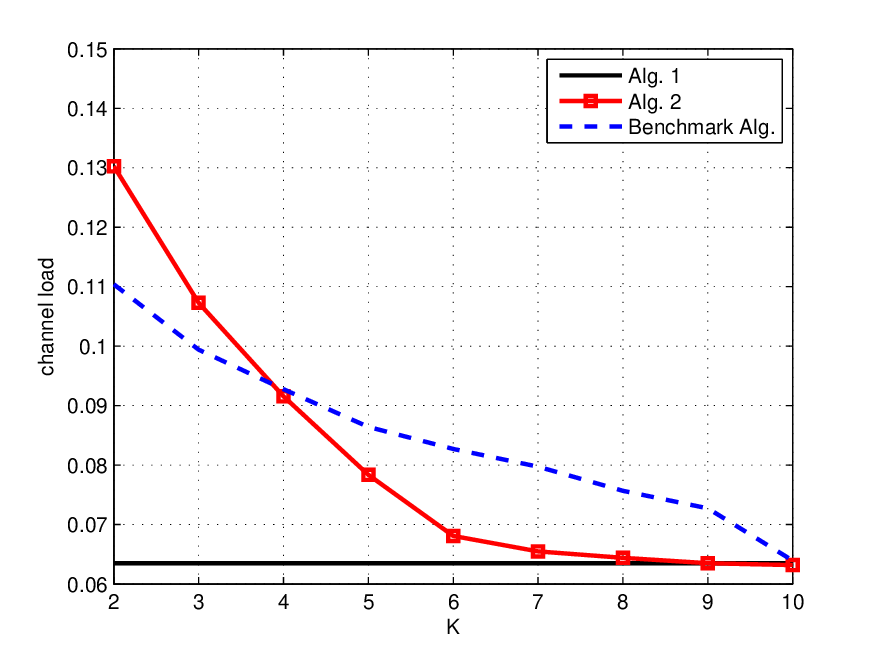}
    \caption{Performance of Algorithm 2 with different values of $K$}
    \label{fig:fig3}
\end{figure}
In Figure 3, the benchmark algorithm achieves better performance
when $K < 4$. However when $K > 4$, Algorithm 2 outperforms
Algorithm 1, which indicates that if $K>4$ the tracking capability
of GPR is sufficient to achieve better performance than that of the
benchmark algorithm.

Figure 4  depicts the average error in the instantaneous average
load estimation when $w=2$, $w=3$ and $w=4$. Clearly, as $K$
increases, the estimation error decreases since the channels are
tracked more accurately for higher values of $K$. This experiment
indicates that the minimum estimation error is achieved when $w=2$.
We conjecture that in this scenario, the channel load measured before the last two
measurements is outdated, and it is more beneficial to use the most
recent measurement results so that the estimation accuracy of GPR is
improved.
\begin{figure}[t]
    \centering
    \includegraphics[width=1.0\columnwidth]{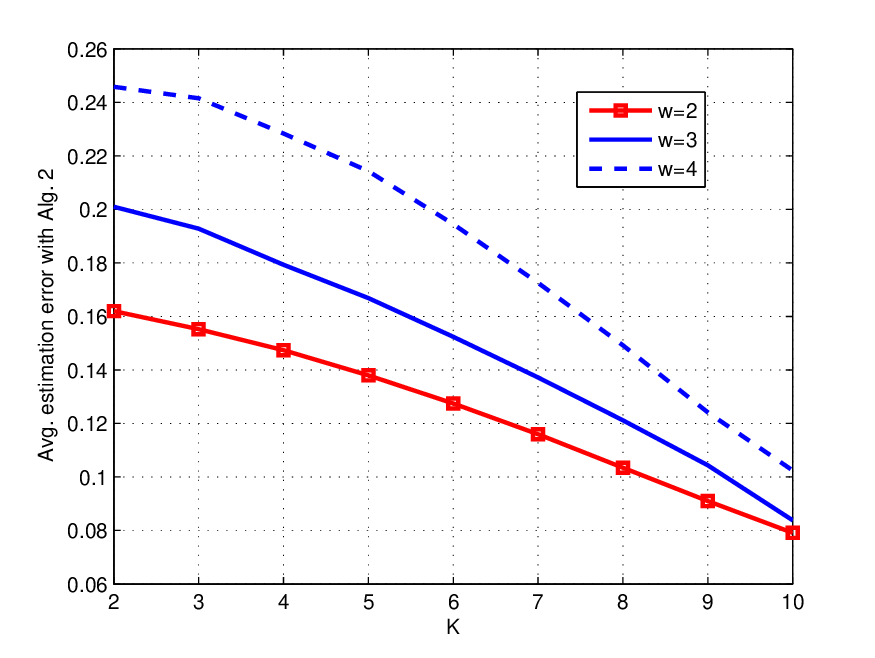}
    \caption{Avg. estimation error with Alg. 2}
    \label{fig:fig4}
\end{figure}

%Lastly, Figure 4 depicts the effects of the number of previous
%measurement data on the performance of Algorithm 2. We increase $w$
%from $w=2$ to $w=7$ for different values of $K$. It can be seen from
%Figure 4 that as $w$ increases  up to 8 samples $L_2$ decreases,
%which indicates that the estimation accuracy of GPR becomes more
%accurate. However, when $w$ is further increased, the accuracy is no
%longer improved. Note that  $w$ can have a significant impact on the
%performance of Algorithm 2 and, must be taken into account as an
%important parameter for algorithm design.
%\begin{figure}[t]
%    \centering
%    \includegraphics[width=1.0\columnwidth]{effectofw.eps}
%    \caption{Effect of $w$ on Algorithm 2 with different values of $K$}
%    \label{fig:fig1}
%\end{figure}
\section{Conclusion }
\label{sec:conclusion}  We have developed a learning based dynamic
channel selection algorithm for the  802.11k supported WLANs. The
proposed algorithm has been designed for the channel measurement
model where only a limited number of channels are allowed to be
monitored at each measurement period. The proposed algorithm first
decides on a set of channels that must be monitored, the selection
of the operating channel is determined by taking into account the
estimated measurement data and the uncertainty levels of each
estimation. We apply GPR to predict traffic load on each
channel based on previous load measurements. In simulation results,
we show that by applying GPR with the proposed algorithm, the cost
associated with channel monitoring can be reduced significantly with
while only causing a small degradation in performance. In this paper, we
assumed that Algorithm 2 takes into account only the uncertainty for
channel selection. More efficient algorithms can be developed by
both considering the estimated load and the uncertainty, which is
our next direction.

\bibliographystyle{IEEEtran}
\bibliography{IEEEabrv,ref}

\end{document}